\documentclass[pra,showpacs,twocolumn]{revtex4}
\usepackage{amsmath}
\usepackage{bm}
\usepackage[dvips]{graphicx}
\usepackage{epsfig}
\usepackage{float}
\usepackage{color}

\begin{document}

\title{Brownian motion theory of the  two-dimensional quantum vortex gas}

\author{Hiroshi Kuratsuji} 

\address{Office of professor emeritus, Ritsumeikan University-BKC, Kusatsu City 525-8577, Japan}

\date{\today}

\begin{abstract}

A theory of  Brownian motion is presented for an assembly of vortices. The attempt  is motivated by a realization of Dyson's Coulomb gas in 
the context of  quantum  condensates. 
 By starting with the time-dependent Landau-Ginzburg (LG) theory, the dynamics of the vortex gas is constructed, which is governed by the canonical 
 equation of motion.  The dynamics of  point vortices  is converted to the Langevin equation, which results in the generalized Fokker-Planck (GFP) 
 (or Smolkovski ) equation using the functional integral on the ansatz of the Gaussian white noise. 
The GFP, which possesses a non-Hermitian property,  is characterized  by  two regimes called the {\it overdamping}  
and the {\it underdamping} regime. 
 In the overdamping regime,  where the dissipation is much larger that  the vortex strength, the GFP 
 becomes the standard Fokker-Planck equation, which is transformed 
into the two-dimensional many particle system.  Several specific applications are given of   the Fokker-Planck equation. An asymptotic limit of small diffusion is also 
discussed for the two-vortices system. The underdamping limit, for which the vortex charge is much larger than the dissipation, is briefly discussed. 
\end{abstract}

\maketitle

\section{Introduction}

The study of vortex gas is one of the main topics in physics. The original idea  dates back to 
the work of Kirchhoff (see  \cite{Lamb}), which describes the dynamics of  an assembly of point  vortices as  
an aggregate of particles  interacting with the logarithmic potential. This idea was transferred to the Onsager's theory 
of vortex gas \cite{Onsager}  aiming  at   exploring  the turbulence. Since then, many  investigations began to 
flourish in  the study of a quantum vortex; see e.g. \cite{Feynman,Fetter,Chiao}. The topics are  still in the 
limelight in condensed matter physics. 

The modern quantum condensed matter physics is based on the concept of 
the order parameter, or more specifically, the macro-wave function  \cite{LP}. The macro-wave function 
provides a useful device to incorporate the vortex degree of freedom in theory. 
In this way,  it is natural to teat the quantized vortex  in the framework of the 
Landau-Ginzburg theory for the order parameter of  the quantum fluids such as superfluids, superconductivity and 
Bose-Einstein condensates \cite{PiSt}.

 In treating condensed matter systems, it is inevitable to deal with the effect of "noise" or "fluctuations." The systematic study of 
 the fluctuation has been one of the central subjects in 
  nonequilibrium  statistical physics \cite{Chandra,Uhlenbeck,Kubo,Kadanoff}.  The recent trend of studying random fluctuation has been focused on Brownian 
motion,  which  is facilitated by various physical situations (see, e.g. \cite{Hanggi}).  As for the quantum condensates, 
the most ready case is caused by the fluctuation of temperature as well as  the effect arising rom existence of  impurities. 
 The effect of random noise has been investigated for the  superconductivity vortices (see. e.g.,  \cite{Dorsey,Enomoto,Feigelman}) 
 and in  a classic article on two-dimensional superfluid vortices  \cite{Ambegaokar}.

In the present article, we address  a Brownian motion model for a quantum vortex gas.  The attempt  is inspired by the 
Dyson theory of  Coulomb gas\cite{Dyson}, which was studied  in connection with the random matrix theory. We take it 
up   from a renewed viewpoint as a problem of  real vortices  in quantum condensates. The starting point  is the canonical equation of motion (Kirchhoff equation)  
for the vortex center, which is facilitated by the time-dependent 
LG theory:  Namely, we  use  the complex order parameter such that it  incorporates  the coordinate vortex center. The resultant 
Hamiltonian is  given by a sum of the logarithmic potential and the harmonic confinement potential. 
The basic  idea is to construct the Langevin equation by modifying the canonical equation of motion for vortex so as to 
include  the dissipation and random force. This is converted to  the Fokker-Planck (FP) equation, 
by adopting the functional integral based on the Gaussian white noise for the random force. The 
procedure follows a  previous article\cite{TK}, which developed a general formulation of the 
stochastic theory of the Schroedinger equation by refining the formalism so as to adapt 
the present purpose. The FP theory indeed provides  a standard framework of the study of random systems (see, e.g. \cite{Risken,Talkner}). 

The crux of the present attempt is as follows: The resultant FP equation is regarded as a  generalization of 
the conventional FP equation [hence   we call it a generalized FP equation (GFP)]. The GFP 
is characterized by three parameters; the diffusion constant,  the dissipation constant and the vortex charge. 
Keeping the diffusion constant 
fixed, there are two cases which depend on the choice of  two constants the dissipation (say, $ \mu $) and the vortex strength  ($ \kappa $), 
namely,   (1) $ \mu \gg \kappa $ and 
(2); $ \mu \ll \kappa $, which is called the "overdamping" and "underdamping" respectively. 
In the latter case  the GFP is written as a form of  the  Liouville equation which can be connected to a transport equation. 

Our interest is mainly focused on  the case of the overdamping,  for which the  GFP is described by 
the  non-Hermitian Schr\"odinger equation resulting in   the standard  FP(or Smolkowski) equation using the "change of phase".   
 On the basis of the  general formulation, we address several  aspects of the FP equation as well as an 
 asymptotic behavior for  the functional integral.  
 
 The content of the paper is as follows: The next section gives a brief sketch of the dynamics of the quantum vortex. 
 In Sec.III a general  framework of the Brownian  theory is developed. In Sec.IV specific problems are  discussed for the FP equation. 
 Sec V is devoted to the "semiclassical analysis" for the functional integral in the small diffusion limit. 
  In the last section we give a brief discussion of  the aspect of the underdamping aspect of the generalized FP equation.

\section{Dynamics of the quantum vortices}

\subsection{Preliminary} 
We start with a brief sketch for the dynamics of the assembly of vortex gas occurring in the quantum boson fluids and spin system
\cite{HK,OK}. Here we have in mind mainly  the Bose-Einstein condensate of atomic gas \cite{PiSt}.  

To begin, let us consider  the complex order parameter written in a polar form: that is, expressed 
in terms of the density $ \rho $ and the phase $ \phi $:  
\begin{equation}
\psi = \sqrt{\rho} \exp[-i\phi] . 
 \end{equation}
The LG Lagrangian density is written as the sum of two terms 
 \begin{align}
 l   &= l_C - l_H \nonumber \\
 &   \equiv  i\hbar (\psi^{*} {\dot\psi} - c.c.) - H(\psi, \psi^{*})  
 \end{align}
 The first term is called {\it canonical term} for convention, whereas 
 the second term represents the Hamiltonian that consists of the 
kinetic energy as well as the interaction terms:
\begin{equation}
H = \frac{\hbar^2}{2m}\nabla\psi^{*}\nabla\psi + V 
\end{equation}
with $ m $ being the mass of constituent particle of condensate. Thus 
the Lagrangian becomes 
\begin{align}
 l  =   \rho \dot\phi - H,  ~~H =\frac{1}{2} m\rho{\bf v}^2 +  V(\rho ),  
\end{align}
where  $ {\bf v} = \frac{\hbar}{m}\nabla \phi $ gives the velocity field.  We restrict the argument to the case  
the potential term is expressed as a function of the density. \\

\subsection{Hamiltonian dynamics of planer vortices}

The first problem is to construct a dynamical equation for the vortex center \cite{HK}. 
In the following argument, we consider  the vortex in  two-dimensional plane $ (x,y) $. 
 We derive the equation of motion for the center of vortices which are 
 parametrized by  $ {\bf R}_i(t) = (X_i(t), Y_i(t)) $ ( $ i = 1 \cdots n $).  
To carry out this we prepare a profile for the density $ \rho $ such that the vortex configuration is incorporated. 
 The field argument $ {\bf x} $  is thus replaced by $ {\bf x} \rightarrow {\bf x} -
{\bf R}_i(t) $.  The phase angle $ \phi = \tan^{-1}(y/x) $ is written in an extended form: 
\begin{equation}
\phi =  \sum_i \kappa_i \tan^{-1}(\frac{y-Y_i(t)}{x-X_i(t)}) \equiv \sum_i \kappa_i \phi_i 
\end{equation}
where the winding number is chosen to be $ n = 1 $.  The attached parameter 
$ \kappa_i $ represents the {\it vortex charge} with $ \vert \kappa_i \vert = \kappa \equiv \frac{\hbar}{m}   $ 
(namely, the same absolute charge for all vortices). 
Using the chain rule $ 
\frac{\partial \phi}{\partial t} = \dot X\cdot \frac{\partial\phi}{\partial X} $ it follows that 
\begin{equation}
L_C = m\sum_i \int \rho \kappa_i {\bf v}_i \cdot \dot{\bf R}_i d^2x 
\label{lc} 
\end{equation}
with  $ {\bf v}_i = \kappa_i \nabla\phi_i  $, the velocity from the i-th vortex, in other words, 
$ m\rho {\bf v}_i  \equiv {\bf p}_i  $ defines the  canonical conjugate to  $ {\bf R}_i  $. 
In what follows we write the effective Lagrangian for the 
assembly of vortices  in terms of the vortex center coordinates $ (X_i(t), Y_i(t)) $. \\

{\it Inter-vortex interaction}: Noting the relation  $ \nabla\phi_i = \kappa_i{\bf k}\times \nabla \log \vert {\bf x}-{\bf R}_i(t) \vert $, 
[$ {\bf k} $ denotes the unit vector in the z-direction that is perpendicular to the 
$ (x,y) $ plane],  $ H $ turns out to be 
\begin{equation}
H =  \sum_{ij}  \kappa_i\kappa_j \int \rho  \nabla \log \vert {\bf x}-{\bf R}_i(t) \vert 
\cdot \nabla \log \vert {\bf x}-{\bf R}_j(t) \vert d^2x . 
\label{int}
\end{equation}
The profile of  density near a vortex is assumed to have a  form:  $ \rho(r) =  \rho_0f(r) $, where 
$ f(r) $ is a monotonically increasing function  such that $ f(0) = 0, ~~f(\infty) = \rho_0 $.  
The integral in (\ref{int}) is carried out for a pair of vortices the center of which are $ {\bf R}_i $ and $ {\bf R}_j $. 
Let us write $ {\bf y} = {\bf x} - {\bf R}_{ij} $ with $ {\bf R}_{ij} = {\bf R}_i -  {\bf R}_j $. By partial integration we have 
$$   - \int \nabla(\rho\nabla \log \vert {\bf y} \vert) \cdot  \log \vert {\bf y} - {\bf R}_{ij} \vert  d^2y 
$$
and noting $ \log \vert {\bf y} - {\bf R}_{ij} \vert \simeq \log \vert {\bf R}_{ij} \vert $ \cite{foot1}, the integral turns out to be 
$ - \log \vert {\bf R}_{ij} \vert  \int^{\infty}_o \frac{d\rho}{dr} dr = \rho_0 $. Thus we arrive at the  familiar form of 
the effective Hamiltonian for the assembly of vortices: 
\begin{equation}
H_{int} = -\frac{1}{2}m\rho_0 \sum_{ij} \kappa_i\kappa_j \log \vert {\bf R}_i - {\bf R}_j\vert . 
\label{log}
\end{equation}
\\

{\it Confinement potential}:  Besides this well known log-potential, we have another term coming from the 
common confinement potential \cite{PiSt}, which  is written as 
$$  H_u  \sim 
\int \Psi({\bf x}, \{{\bf R}_i\})^{*}\Psi({\bf x}, \{{\bf R}_i\}) U({\bf x}) d^2x . 
$$
Taking account of the vortex profile, one sees that  $ \Psi^{*}({\bf x})\Psi({\bf x}) \sim 
\sum_i \rho({\bf x} -{\bf R}_i ) $. Then the contribution from each vortex can be 
approximated by the well localized function. To obtain the actual value, we need to subtract 
the contribution from the uniform background $ \rho = \rho_0 $. 
The profile of the function $ f(r) $ can  be chosen as the Gaussian form  
$ f(r) = \exp[ -ar^2]  $,  therefore we get the contribution from the confinement  potential 
\begin{equation}
H_u \sim  \int \rho_0 U({\bf x})d^2x 
 - \int \rho_0 (1- f({\bf x} - {\bf R})) U({\bf x}) d^2x  = U({\bf R}).  
\end{equation}
As the most  typical case, we adopt the 
harmonic oscillator form: $ U = k{\bf x}^2 $.  So the contribution from $ n $ vortices is given by $ k\sum_{i=1}^n {\bf R}_i^2 $. 
By summarizing  the log potential and the confinement potential term, we have the effective Hamiltonian 
\begin{equation}
H_{eff} =  -\frac{1}{2}m\rho_0 \sum_{ij} \kappa_i\kappa_j \log \vert {\bf R}_i - {\bf R}_j\vert 
+ \sum_i k{\bf R}_i^2 . 
\end{equation}
Here we note that there are additional contributions arising  from 
the {\it pinning} effect  caused by the presence of impurities, which we do not use explicitly in the following argument. \\

{\it Canonical equation of motion}: We turn to the canonical term $ L_C $, (the first term of the Lagrangian). 
Here instead of calculating  $ L_C $ directly in terms of the vortex center coordinate $ {\bf R} $, 
we derive  the equation of motion by using the differentiation under the integral symbol. After some manipulation, 
we obtain 
\begin{eqnarray}
{\bf F}_{ix}^C 
 & = &  {d \over dt}{\partial L_C \over \partial \dot X_i}
            - {\partial L_C \over \partial X_i}  \nonumber \\
 & = & m\int \big\{\frac{\partial (\rho v_{iy})}{\partial x} -    \frac{\partial(\rho v_{ix})}{\partial y} \big\} d^2x . 
\end{eqnarray} 
Here the integral becomes the line integral $ \int \rho {\bf v}\cdot d{\bf s} $ which yields $ 2\pi \rho_0 $ by taking account of the boundary condition 
$ \rho(\infty) = \rho_0 $,  hence we get $ F^C_{ix} = 2\pi m\rho_0 \dot Y_i $.  
In a similar way, it follows that $ F_{iy}^C = - m\rho_0 \dot X_i $. Writing these in terms of the vector notation, we have 
\begin{eqnarray}
{\bf F}_i^C 
 =  {d \over dt}{\partial L_C \over \partial \dot{\bf R}_i}
            - {\partial L_C \over \partial {\bf R}_i} 
  =    2\pi m\rho_0(\kappa_i {\bf k} \times \dot{\bf R}_i) . 
\label{x}            
\end{eqnarray}
 From the above result, one  can guess the form of the Lagrangian, which is simply given as 
\begin{equation}
L_C =  2m\rho_0 \sum_i\kappa_i (Y_i\dot X_i - X_i\dot Y_i) . 
\end{equation}

Thus the effective Lagrangian including the effective Hamiltonian 
is given by $ L_{eff} = L_C - H_{eff} $.  In the following argument we focus the argument on the special case that 
  $ \kappa_i = \kappa $ for all indices $ i $; namely, all vortices have the same charge. 
In this way the equation of motion for the assembly of  vortices becomes 
\begin{equation}
\bar\kappa 
\frac{d X_i}{dt} 
 =  \frac{\partial H_{eff}}{\partial Y_i}, 
~~\bar\kappa \frac{d Y_i}{dt}  
=  -\frac{\partial H_{eff}}{\partial X_i}
\label{canonical}, 
\end{equation}
where we use an abbreviated symbol:  $ \bar\kappa \equiv 2m\rho_0\kappa $. 
This set of equations of motion implies that  the pairs $ (X_i, Y_i) $ form canonical variable each other. 
Alternatively, this is written in terms of  vector notation: 
\begin{equation}
\bar\kappa {\bf k} \times 
\frac{d {\bf R}_i}{dt} 
 =  \frac{\partial H_{eff}}{\partial {\bf R}_i}, 
\label{canonical1} 
\end{equation}
This vector equation represents  the balance between two type forces: the left hand side represents the "Magnus force" and the 
right hand side is the potential force. 

Thus the  present vortex Hamiltonian has  just the same form as the one for the charged particles interacting through the 
Coulomb repulsion confined in the harmonic potential. \\

{\it An example of two point-vortex}:  
For this simplest problem the Hamiltonian becomes 
\begin{equation}
H_{eff} = -\frac{1}{2}m\rho_0\kappa^2 \log \vert {\bf R}_1 - {\bf R}_2 \vert + \frac{1}{2}k({\bf R}_1^2 + {\bf R}_2^2) . 
\label{twovortex}
\end{equation}
This can be separated into the relative and center-of-mass coordinates, 
$$
 2{\bf r} = {\bf R}_1 - {\bf R}_2, ~~ 2{\bf R}_G =  {\bf R}_1 +  {\bf R}_2, 
$$
by which the $ H_{eff} $ is given by the sum: 
\begin{equation}
H_r = -\frac{1}{2}m\rho_0\kappa^2 \log \vert {\bf r} \vert + k{\bf r}^2,  ~~H_G = k{\bf R}_G^2 .   
\label{rRg}
\end{equation}
Thus the equation of motion turns out to be 
\begin{equation}
\frac{d}{dt} ( {\bf r} ^2 ) =0, ~~\frac{d}{dt} ( {\bf R}_G^2 ) =0, 
\end{equation}
which results in  the energy surfaces: $ H_r = E_r, H_G = E_G $. 
 This case of  two- vortices will be used for the solution for the FP equation in later section.

 \section{The Brownian motion }

  \subsection{The Langevin equation}
  
Now we discuss the stochastic aspect on the basis of the vortex dynamics  given above. 
The effect of fluctuations arises from a variety of origins mainly caused by impurities. 
If these impurities are distributed in an irregular way, 
the vortex centers will acquire randomness as a result of, for example, the effect of scattering. 
Apart from such scattering effect among vortices, the thermal effect may provide the more 
direct effect. 
Then, it is natural to expect that the Brownian motion of the vortex centers occurs, 
which can be describe by a random fluctuation denoted  as $ {\bf{\xi}} $.

Besides the random fluctuation, we have to take account of the effect coming from dissipation,
which is required from the {\it fluctuation-dissipation theorem} (see e.g.\cite{Kubo}). 
Indeed the dissipation can be caused by the inevitable effect of absorption of the vortex energy caused 
by an interaction with environment, more specifically the dissipation arises from the interaction with the normal fluid 
component (see, e.g.\cite{Ambegaokar}). 

Here we mention an early attempt of the Langevin and FP approach of the quantum vortex that was 
worked out  in \cite{Ambegaokar}. 
This paper gave a dynamical theory of  the vortex pairs of opposite charge aiming at a description of 
dissociation of a bound pair of vortices. On the other hand, in 
the present attempt we are concerned with a quite different aspect of a two dimensional vortex gas,  namely, 
the repulsive Coulomb gas in the confinement potential.

 To be noted here is that  the random force  $ \xi_i $
acting on the $ i  $ -th vortex is independent of $ \xi_j $, that is, these have  no correlation with each other. 
 Furthermore, the dissipative force is common to all 
the vortices, which is denoted by $ \mu $. Indeed this assumption on the fluctuation and dissipation 
 is the simplest  and most reasonable. 

Now  taking account both dissipation and random noise, the gradient force in Eq.~(\ref{canonical1}) can be  simply modified such that 
$$
  \frac{ \partial H_{eff} }{ \partial {\bf{R}}_i } \rightarrow \frac{ \partial H_{eff} }{ \partial {\bf{R}}_i } + \mu \frac{ d {\bf{R}}_i }{ dt } + \xi_i. 
$$
 This modification is somehow a well known procedure in general condensed  matter physics. We here borrow  the procedure used for 
  ferromagnetic particles  \cite{Brown} which deals with the thermal fluctuation of the magnetic moment. In the present case the 
 magnetic moment is replaced by the velocity of the vortex center $ \dot{{\bf R}} $.  
  By taking account of this prescription,  the equation of motion is given  as follows:
\begin{eqnarray}
   \bar\kappa ({\bf{k}} \times \frac{d{\bf{R}}_i }{ dt }) = \frac{ {\partial} H_{eff} }{ {\partial} {\bf{R}}_i } + \mu \frac{ d {\bf{R}}_i }{ dt } + \xi_i . 
  \label{canonical2}
\end{eqnarray}
Multiplying $ {\bf k} $ by  (\ref{canonical2}) one obtains 
\begin{eqnarray}
  \bar\kappa \frac{ d {\bf{R}}_i}{ dt } = - {\bf{R}}_i \times \left( \frac{ {\partial} H_{eff} }{ {\partial} {\bf{R}}_i } 
  + \mu \frac{ d {\bf{R}}_i }{ dt } + \xi_i\right) . 
  \label{canonical3}
\end{eqnarray}
 Using these two equations (\ref{canonical2}) and (\ref{canonical3}), the equation for $ \dot {\bf R}_i $ can be derived \cite{note}:  
\begin{eqnarray}
  \frac{d{\bf R}_i}{dt} = - {\bf{A}}_i + {\bf{\zeta}}_i . 
  \label{langevin}
\end{eqnarray}
where we adopt the following scaling of  time variable $ t \rightarrow (\mu^2 + \bar\kappa^2)t $  \cite{foot2}
and  ${\bf{A}}_i$ and $\zeta_i$ are given  as  
\begin{eqnarray}
{\bf A}_i & =  & 
 \left[ {\mu} \frac{ {\partial} H_{eff} }{ {\partial} {\bf{R}}_i } 
  + \left( \bar\kappa{\bf k} \times \frac{ {\partial} H_{eff} }{ {\partial} {\bf{R}}_i } \right) \right] ,  \label{Avector} \\
  \zeta_i & = & - \left[ {\mu} {\xi}_i + \bar\kappa  {\bf k} \times {\xi}_i \right] . 
\label{fluctuation}
\end{eqnarray}
The expression (\ref{fluctuation})  can  be written in terms of $ (x,y) $ component: 
\begin{equation}
\zeta_i^x = \mu\xi_i^x -  \bar\kappa\xi_i^y, ~~\zeta_i^y 
 = \bar\kappa \xi_i^x +  \mu \xi_i^y ,  
\label{fluctuation2}
\end{equation}
which means the orthogonal transformation in two dimensional plane with the rotational angle;
 $ \tan \Theta = \frac{\bar\kappa}{\mu} $ 
 ,namely,(\ref{fluctuation2}) represents 
{\it just}  a rotation of the original fluctuation $ \xi $, so that  the "additive  nature" is safely kept.

Without  $ \zeta $, Eq.~(\ref{langevin}) is an analogy of the ``Landau-Lifschitz equation'',
which  is well known in ferromagnetic theory \cite{Brown,LL2}.  
We note that the vector $ {\bf A}_i  $ consists of two terms: the first term can  be called 
the {\it gradient term} , and  the second term {\it gyration term}, which can  be a counterpart of the "Magnus force". 
The mutual interplay between these two terms 
characterizes the stochastic process of the vortex motion,  which we discuss below. 
 
Now we put an Ansatz of  the Gaussian white noise for $ \xi_i $ , which satisfies the 
correlation: 
\begin{eqnarray}
\langle \xi_i^{\alpha}(t) \rangle & = & 0 \nonumber \\
  \langle \xi_i^{\alpha}(t)\xi_{j}^{\beta}(t+u) \rangle & = & h \delta_{\alpha\beta}\delta_{ij} \delta(u), 
  \label{gwn}
\end{eqnarray}
where the suffix  $ (\alpha, \beta) $ represents $ (x,y ) $ and $ \delta(u) $ means the delta function. 
The diffusion constant $ h $  is assumed to take a common value for 
all  components $ i $.  The correlation (\ref{gwn}) is transferred to the resultant random force  (\ref{fluctuation}) $ \zeta = (\zeta^x, \zeta^y) $; 
we have  that the same form of the correlation  
\begin{eqnarray}
\langle \zeta_i^{\alpha}(t)\rangle & = & 0,  \nonumber \\
  \langle \zeta_i^{\alpha}(t)\zeta_{j}^{\beta}(t+u) \rangle &  = &  (\bar\kappa^2 + \mu^2)h \delta_{\alpha\beta}\delta_{ij} \delta(u) . 
\label{GWN}
\end{eqnarray}
Taking into account this feature, it is possible  to replace  $ (\bar\kappa^2 + \mu^2)h  \rightarrow h $ in the following argument.  
Thus the above set of the Langevin equations obeys a set of the random forces 
$ \zeta_i: \{i = 1 \sim n\} $ which satisfy  the common correlation relation independent of the 
vortex indices.

\subsection{The functional integral and non-Hermitian Schroedinger equation}

If we recall  that  constituent vortices  are  independent of each other,  the probability distribution 
 for an assembly of  vortices is given by the product of the Gaussian noise for  $ ({\zeta}_i : i= 1 \sim n)$, which 
 becomes the standard Gaussian functional form \cite{Fibich}: 
\begin{equation}
 P[\{\zeta_i(t)\}] = \prod_{i=1}^{n}\exp \left[- \frac{1}{2h} \int_0^t  {\bf \zeta}_i^2(t) dt \right] . 
 \label{GF}
 \end{equation}
Using this distribution, the transition probability
from $ {\bf{R}} (0) $ to $ {\bf{R}} (t) $ is given by the path integral
\begin{equation}
 K[{\bf R}(t)\vert {\bf R}(0)] = \int_{ {\bf{R}} (0) }^{ {\bf{R}} (t) }   \prod_i \exp \left[- \int^{t}_{0} \frac{{\bf \zeta}_i^2(t)}{2h}dt \right]
  \mathcal{D}[{\bf \zeta}_i(t)] , 
  \label{transition}
\end{equation}
where the notation for a set of vortex centers  $ {\bf R} \equiv ({\bf R}_1 \cdots {\bf R}_n) $ is used. With  this expression, the 
process of transition caused by the random noise can be built in an implicit way.  

In order to explicate the process of building up the path integral over the orbits in the space  of vortex center $ {\bf R} $, 
we adopt the following steps: To ensure the Langevin equation, we insert the expression of the $\delta$--functional integral
\begin{equation}
  \int  \prod_{i=1}^n \prod_t\delta[ {\bf F}_i(t) - \zeta_i(t)] \mathcal{D}{\bf F}_i(t) = 1 , 
\label{delta}
\end{equation}
where we use the notation 
\begin{equation}
  {\bf F}_i =  \frac{d{\bf R}_i}{dt} + {\bf A}_i 
  \end{equation}
  and $ \prod_t $ means the {\it continuous  product} over time interval $ [0,t] $.  
  Using the above delta functional, the transition rate (\ref{transition}) can be brought to the path integral
\begin{align}
 K[{\bf R}(t)\vert {\bf R}(0)] &  = \int_{ {\bf{R}} (0) }^{ {\bf{R}} (t) } 
 \prod_i \exp \left[- \int^{t}_{0} \frac{{\bf \zeta}_i^2(t)}{2h}dt \right] \nonumber \\
  &  \times \prod_{i=1}^n \prod_t\delta[ {\bf F}_i(t) - \zeta_i(t)] \mathcal{D}{\bf F}_i(t)  \mathcal{D}[{\bf \zeta}_i(t)] , 
  \label{transition2}
\end{align}
 which enables us to derive the FP equation in most direct way. The intermediate step 
  will be given in Appendix A.  Noting this result,   we use an  ``imaginary time trick'', that is,
we define $ {\tau} = - i t $, and then the propagator is written in the {\it quantum mechanical }
path integral form 
\begin{equation}
 K[{\bf R}(\tau)\vert {\bf R}(0)] = \int \exp \left[ \frac{i}{h}\int  {\cal L} d\tau \right] \mathcal{D} [{\bf R}]. 
\label{PI}
\end{equation}
Here the {\it Lagrangian } becomes 
\begin{equation}
 {\cal L} = \sum_i\big\{\frac{1}{2}\left(\frac{d{\bf R}_i}{d\tau}\right)^2 + i{\bf A}_i\cdot \frac{d{\bf R}_i}{d\tau} \big\} - W
\end{equation}
with the potential function 
\begin{equation}
  W =  \sum_i (\frac{ {\bf A}_i^{2} }{ 2 } -  M_i h),  
  ~~M_i  =  \frac{1}{2} \frac{ \partial }{ \partial {\bf{R} }_i} \cdot {\bf{A}}_i , 
\end{equation}
where the second term in $ W $   comes from the Jacobian written in an imaginary time form
(see Appendix A): 
\begin{align}
J & = \exp \left[ \frac{i}{h}  \int_{0}^{\tau} \sum_i M_ih d\tau \right] . 
\end{align}

Now by introducing  the ``wave function'' $ \Psi({\bf{R}}, \tau) $, we write the integral equation:
\begin{equation}
  \Psi({\bf R}, \tau) = \int K[{\bf R}(\tau)\vert {\bf R}(0)] \Psi({\bf R}, 0) d{\bf R}(0) . 
\end{equation}
Following the standard procedure of Feynman path integral, we obtain the "Schr{\"o}dinger equation"~\cite{Schulman}:
\begin{eqnarray}
  i h \frac{ {\partial} {\Psi}}{ {\partial} {\tau} } = \frac{1}{2} \sum_i \left( {\bf{p}}_i  - i {\bf{A}} _i\right)^{2} \Psi + W \Psi , 
  \label{complexFP}
\end{eqnarray}
where $ {\bf{p}} = - i h \frac{ {\partial} }{ {\partial} {\bf{R}} } $.   This form of the wave equation has the same form as 
the particle in the presence of the {\it vector potential} $ {\bf A} $.  
We note that  (\ref{complexFP}) is apparently  non-Hermitian in general  (see e.g., \cite{NH}); more details on this point will be discussed later) . 

By replacing the imaginary time $ {\tau} $ with the original (genuine) time  $ t $, namely $ {\tau} \rightarrow - i t $, and 
rewriting the wave function $ \Psi $ by the distribution function $ P $, then one arrives at  
\begin{align}
 \frac{\partial P }{ \partial  t } & = \frac{h}{2}\sum_i \frac{ \partial^2P  }{ \partial{\bf R}_i^2}, 
\nonumber \\
   &  + \sum_i   \frac{ {\partial} }{ {\partial} {\bf{R}}_i }
   \cdot \left( 
\big[ \mu \frac{\partial H_{eff}}{\partial {\bf R}_i} + ( \bar\kappa{\bf k}\times \frac{\partial H_{eff}}{\partial {\bf R}_i})\big]  P \right) . 
\label{GFP}
 \end{align} 
Here use is made of the relation  $ ({\nabla}\cdot {\bf A})P + {\bf A}\cdot \nabla P = \nabla \cdot({\bf A}P) $. 
Equation (\ref{GFP}) is the main consequence of the present paper, though it looks simple enough. 
This can be  regarded as a two-dimensional generalization of the  FP(Smolkowski) equation 
used in Dyson's theory\cite{Dyson}, where the  second term in $ {\bf A} $, namely, the gyration term  is missing.  
We rewrite  this generalized FP equation  in the form of  the conservation of current: 
\begin{equation}
  \frac{ {\partial} P }{ {\partial} t} + \nabla\cdot  {\bf j} = 0
\end{equation}
with $ \nabla \equiv \sum_{i} \frac{\partial}{\partial {\bf R}_i} $.  The probability current is defined as the sum
\begin{align}
{\bf j} & = {\bf j}_1 + {\bf j}_2 ,  \nonumber \\
{\bf j}_1 & = - \frac{h}{2 }\nabla P - \mu \sum_i\frac{\partial H_{eff}}{\partial {\bf R}_i}P \\
{\bf j}_2 & = {\bf k} \times \big[ - \frac{h}{2 }\nabla P -  \bar\kappa\sum_i\frac{\partial H_{eff}}{\partial {\bf R}_i}P\big] . 
\end{align}
These  expressions are  significant; the first term of the respective terms stand for a diffusion effect,
and the second term represents the ``transport of probability mass, '' 
namely, $ \sum_i \frac{ d {\bf{R}}_i }{ dt } P $. 
This feature suggests that the FP equation thus obtained
describes a diffusive behavior of the vortex gas. 

Now we recall  that the vector $ {\bf A} $ consists of two terms; the gradient and the gyration terms.  These two terms are controlled 
by a  competition between two parameters, $ \bar\kappa $  and $ \mu $,  the vortex charge  and the magnitude of 
dissipation respectively.   We have two {\it extreme} cases. (1) The relation $ \mu \gg \kappa $ holds, that is, the gradient term becomes dominant, 
and the gyration term discarded in (\ref{GFP}). This case is called  "overdamping". 
 (2) On the other hand,  if $ \mu \ll   \bar\kappa   $, this corresponds to "underdamping," for which 
 the gyration is dominant.  The original idea to separate these cases dates back to  the articles \cite{Langer,Landauer}.  

 In what follows,  we restrict the  argument to the overdamping approximation and the 
 underdamping case will be briefly sketched in the last section.    Thus we obtain 
\begin{align}
 \frac{\partial P }{ \partial  t } & = \frac{h}{2}\sum_i \frac{ \partial^2P  }{ \partial{\bf R}_i^2} 
\nonumber \\
   &  + \sum_i \mu  \frac{ {\partial} }{ {\partial} {\bf{R}}_i }
   \cdot \left( 
\big[\frac{\partial H_{eff}}{\partial {\bf R}_i}\big]  P \right) 
\equiv {\cal L}_{FP} P . 
\label{FP}
 \end{align} 
If (\ref{FP}) is regarded as a non-perturbed term,  the term coming from the gyration  $ {\bf j}_2 $  can  be treated as a perturbation.

\section{Specific aspects of the FP equation}

\subsection{Statistical mechanical  consequences}

{\it  Statistical average using  the FP equation}: We first examine the general properties concerning the FP equation.  
Here we take up a typical example;  the evolution of the statistical average for the function $ \langle K{\bf R}_1,\cdots {\bf R}_n)\rangle $. 
Using the FP equation this satisfies the equation 
\begin{equation}
\mu \frac{d \big\langle K \big\rangle }{dt} = - \sum_i \big\langle \frac{\partial H_{eff}}{\partial {\bf R}_i}\frac{\partial K}{\partial {\bf R}_i} \big\rangle 
+ \frac{h}{2}\sum_i \big\langle \frac{\partial^2 K}{\partial {\bf R}_i^2}\big\rangle . 
\end{equation}
As a special case, we consider the moment of the vortex center: $ K = \sum_i {\bf R}_i^2 $, for which we have 
\begin{equation}
\mu\frac{d \langle K \rangle}{dt} = (\frac{hN}{2} - k^2) - 2k \langle K \rangle , 
\end{equation}
which leads to 
\begin{equation}
\langle K \rangle = K_0 \big(1 - \exp[-\frac{2k}{\mu}t] \big) 
\end{equation}
with $ K_0 = \frac{1}{\mu}\big( \frac{hN}{2} - k^2\big) $. \\

{\it Stationary distribution : 2-dimensional Coulomb gas}:   Let us consider the stationary state, namely distribution function 
satisfies $ \frac{\partial P}{\partial t} = 0 $.  If we put an Ansatz; $ P = \exp[ -\beta H_{eff}] $ and substitute this into (\ref{FP}), 
we get the relation 
 \begin{equation}
   (\frac{h\beta}{2} - \mu)\big[ \sum_i \big\{\frac{\partial^2H_{eff}}{\partial {\bf R}_i^2}
   -   \beta\big(\frac{\partial H_{eff}}{\partial {\bf R}_i}\big)^2\big\}\big] =0 . 
 \end{equation}
From the requirement that this relation should hold for arbitrary Hamiltonian,   we get 
\begin{equation}
  \frac{h\beta}{2}  = \mu , 
  \label{FD}
  \end{equation}
 which is nothing other than the content of the fluctuation-dissipation relation \cite{Kubo}. 
 In this way the equilibrium state of  the assembly of vortices is realized as the two-dimensional Coulomb gas \cite{Dyson}, for which 
 $ P  = \exp[-\beta H_{eff}] $ becomes the Boltzman factor 
\begin{equation}
P  = \exp[ \kappa^2\beta  \sum_{ij} \log \vert {\bf R}_i  - {\bf R}_j \vert ] 
\times \exp[  - \frac{1}{2}k \beta \sum_i \vert {\bf R}_i \vert^2 ] 
\end{equation}
with $ \beta \equiv 1/k_B T  $ being the inverse temperature. 
It is to be mentioned that $ \exp[-\beta H_{eff}] $ is the eigenfunction of $ {\cal L}_{FP}  $ with the 
{\it zero eigenvalue},  namely  $  {\cal L}_{FP} (\exp[-\beta H_{eff}]) = 0 $ (see the argument below).

\subsection{Reduction to a  quantum mechanical many particle system}

As has been noted in the previous section, the "wave equation" (\ref{complexFP}) is  non-Hermitian Schr\"odinger  equation.   
 In the overdamping case, for which the gyration term is discarded, one can eliminate the {\it vector potential } $ {\bf A} $ 
 by adopting the change of phase  \cite{Dirac},  $ \Psi = \exp[ - if ] \tilde\Psi $, where $ f $ is chosen such that the following 
  equation holds 
\begin{equation} 
 \nabla f =  \frac{1}{h} 
\sum_i {\bf A}_i \label{gauge}
\end{equation}
from which one sees that  $ f $ is proportional to $ H_{eff} $; hence noting the relation (\ref{FD}), then we write the 
transformation like 
\begin{equation}
P = \exp[-\frac{\beta H_{eff}}{2}] \tilde P . 
\end{equation}
Thus one can reduce the FP equation to the familiar form of the Schr\"odinger-type  equation for  $ \tilde P $ 
\begin{align}
\frac{\partial \tilde P}{\partial t}  & = \frac{h}{2}\sum_i \nabla_i^2 \tilde P + 
\mu\big[\frac{1}{2}\sum_i \nabla_i^2 H_{eff}   \nonumber \\
 & - \frac{\beta}{4}(\sum_i \nabla_i H_{eff})^2\big]\tilde P   \equiv  \hat{\cal H}_{FP} \tilde P,  \\
 \hat{\cal H}_{FP} & = \exp[\frac{\beta H_{eff}}{2}] 
 {\cal L}_{FP} \exp[-\frac{\beta H_{eff}}{2}], 
 \label{Schrodinger} 
\end{align}
which is explicitly written as 
\begin{align}
\frac{\partial \tilde P}{\partial t}  & = 
\big[\frac{h}{2}\sum_i \nabla_i^2  - V({\bf R}_1, \cdots {\bf R}_N)\big] \tilde P  \equiv  \hat{\cal H}_{FP}\tilde P , 
\nonumber \\
  V & = \mu\big[B\sum_i {\bf R}_i^2  + A\sum_{i,j}  \frac{1}{\vert {\bf R}_i - {\bf R}_j \vert^2}\big], \nonumber \\
  A  = &  \frac{1}{4}(m\rho_0)^2\kappa^4, ~~B = 4k^2 
  \label{inverseR^2}
\end{align}
up to some additional constant. We see that the following equation holds 
$$
\hat{\cal H}_{FP} (\exp[-\frac{\beta H_{eff}}{2}])  = 0 , 
$$
which indicates  that $ \exp[-\frac{\beta H_{eff}}{2}]  \equiv  \tilde P_0 $ is an eigenfunction of $ \hat{\cal H}_{FP} $ 
corresponding to zero eigenvalue.  Here (\ref{inverseR^2})  is a typical two dimensional many-body problem,  
which  had once been studied as a major topics 
(e.g. \cite{tomonaga,BP}), for which a brief sketch will be given in Appendix B. 

\subsection{Special case of  two vortices} 

For this case, as is seen from (\ref{rRg}), the relative and the center of mass coordinate 
 is separated and one can put aside the center of mass degree. Hence we can treat the problem 
in such way that one vortex is fixed at the origin, say  $ {\bf R}_2 $.  So 
let us write the "wave function" in the form  $ \tilde P = \exp[-\epsilon t] G $, then 
it turns out to be 
\begin{equation}
\big[ \frac{h}{2}(\nabla^2 + (\epsilon - V(r)\big]G   = 0 , 
\label{eigen}
\end{equation}
where $ r \equiv \vert {\bf R}_1 \vert $ , hence the potential becomes 
\begin{equation} 
V(r) = Br^2 + \frac{A}{r^2} . 
\end{equation}
Let us put $ G(r,\theta) = v(r)u(\theta) $ and choose the angular part to be constant. 

The time-dependent distribution function (the solution for the reduced FP equation)  is written in the form
\begin{equation}
\tilde P(r,t) = \tilde P_0(r)  + \sum_{n=0}c_n(t) \exp[-\epsilon_n t] G_n(r) . 
\end{equation}
Here we need to separate the zero energy solution from the other parts  and 
the coefficients $ c_n(t) $ are determined by the initial condition $ \tilde P(r,0) $, which will be given below. 
We first  solve the eigenvalue equation (\ref{eigen}), for which we change the eigenfunction: 
$ v(r) = \sqrt{r}u(r)  $, then  (\ref{eigen}) turns out to be  
\begin{equation}
\frac{d^2u}{dr^2} + \frac{2}{r}\frac{du}{dr} + \big\{  \epsilon - (Br^2+ \frac{A}{r^2}\big)\big\}u = 0 , 
\label{u}
\end{equation}
where  $ A $ shifts by an amount $ \frac{1}{4} $,  namely, $ A \rightarrow A - \frac{1}{4} $: 
 The eigenvalue equation is known to be analytically solved \cite{LL}, from which 
  we simply borrow the result: Namely, by putting $ \xi = \sqrt{2B/h}r^2) $  $ u(r) $ satisfies 
 \begin{equation}
 \xi \frac{d^2u}{d\xi^2} + \frac{3}{2}\frac{du}{d\xi} + \big[ n + s + \frac{3}{4} - \frac{\xi}{4} 
  - \frac{s(s+ 1/2)}{\xi} u = 0 , 
 \end{equation}
 where $ s, n $ is settled such that $ 2s(s+1) = \sqrt{B/h} $ and $ \sqrt{B/h} \epsilon = 4(n+s) +3 $. Hence 
the eigenvalue $ \epsilon$ is given by 
\begin{equation} 
\epsilon_n = \sqrt{\frac{hB}{2}}\big[ 4n + \sqrt{\frac{4(A- 1/8)}{h}} \big] 
\end{equation}
with $ n  $ being the non-negative integer.  From this expression $ A \geq 1/8 $ should hold. The corresponding eigenstate 
is given by the hypergeometric function 
\begin{equation}
u_n(\xi) = F(-n, 2s+ \frac{3}{2} , \xi). 
\end{equation}
Hence we obtain the solution up to the lowest " excited state", namely, by choosing only the case 
$ n= 0, 1 $ 
\begin{align} 
\tilde P(r,t) & = \exp[-\frac{\beta H_{eff}}{2}] \{ \tilde P_0 + c_0u_0(r)\exp[-\epsilon_0t]  \nonumber \\
 & + c_1u_1(r) \exp[-\epsilon_1 t] \} . 
\end{align}
In this way, we see that in the limit of $ t \rightarrow \infty $, the equilibrium state 
is recovered: $ P \simeq \exp[ - \beta H_{eff}] $.  The above result can  be used for 
a basis of the more complicated system.

Here a remark is in order concerning the above solution. The problem is connected with the confinement 
harmonic potential that is reflected in the coefficient $ B $. About this point,  we note that it  is 
possible to choose the more general case that 
the confinement potential  has an anisotropic and even letting it allow  time varying. By extending the problem 
to such a general case, there may 
appear  a variety of problems; e.g. if this  anisotropy of the harmonic oscillator changes adiabatically, we 
expect adiabatic control of the stochastic behavior of the vortex system.

\section{Small diffusion limit} 

If  we get back to the starting  functional integral, it is intriguing to examine  an asymptotic limit 
in which the diffusion constant $ h $ is regarded as small. We here look for an alternative way 
to obtain an approximate scheme  so as to approach to random behavior for the vortex gas apart 
from the FP equation. 

We consider a system of two vortices for  which one vortex is pinned at the origin as 
before; hence the functional integral is written in the form: 
\begin{equation}
K =  \int \exp[ -\frac{1}{h} \int {\mathcal L}dt ] {\cal D}(r). 
 \nonumber 
\end{equation}
Here  the ``Lagrangian" is given as 
\begin{equation}
{\mathcal L}  = (\frac{d{\bf r}}{dt} + \mu \frac{\partial H_r}{\partial {\bf r}})^2. 
\end{equation}
Using the polar coordinate, one write  $ \frac{d{\bf r}}{dt}  = \dot r \hat r  + r\dot\theta \hat\theta $. 
($ \hat r $ and $  \hat\theta $ stand for the unit vector of radial and its perpendicular direction). 
In the limit of $ h \simeq 0 $, the functional integral is treated by the stationary phase  approximation, which is written in a form 
$ K_{cl}  = \exp[-\frac{1}{h}S_{cl}] $. Here $ S_{cl} $ denotes the {\it classical action} that satisfies the extreme 
condition $ \delta S = 0 $. The extreme condition leads to the Euler-Lagrange equation. 
We have the contribution from the deviation of extreme path that is  
written in terms of the Gaussian functional integral with respect to the deviation from extreme path. 
However, we discard this for the sake of simplicity.  We note a peculiar feature of the Lagrangian:
The variable $ \theta $ does not appear in the Lagrangian, namely, 
$ \theta $ is cyclic coordinate, so the ``momentum " conjugate to $ \theta  $
 is the constant of motion, which is given by  $ r^2\dot\theta = C $. Thus, following the well known procedure 
in analytical dynamics, we construct the Rouse function \cite{LL1}, in which  the $ \theta  $ variable is eliminated to be 
\begin{equation}
 R = C\dot\theta  - {\mathcal L} 
= - (\dot r + \mu\frac{\partial H_r}{\partial r})^2  + \frac{C^2}{r^2}. 
\label{Rouse}
\end{equation}
Thus the equation of motion is derived using the Euler-Lagrange equation  
\begin{equation}
\frac{d}{dt}\big(\frac{\partial R}{\partial \dot r}\big) - \frac{\partial R}{\partial r}= 0. 
\nonumber 
\end{equation}
By  substituting the solution (classical orbit) of this equation into the expression $ {\mathcal L } $;    
\begin{equation}
K_{cl}   =  \exp[-\frac{1}{h}\int_{t_i}^{t_f} \{(\dot r + \mu\frac{d H_r}{dr})^2 + \frac{C^2}{r^2}\} dt] . 
\end{equation}
We look for a further reduced form of $ K_{cl} $; that is, we consider 
the case that $ C $ is regarded as small enough such that it is  treated as 
perturbation parameter. Hence we can omit the last term in (\ref{Rouse}); so the equation of motion 
becomes in a simple form:  
\begin{equation}  
\dot r + \mu\frac{d H_r}{dr}= 0 , 
\label{instanton}
\end{equation}
which is solved to lead to the orbit: $ (r,\theta) $: 
\begin{equation}
r^2(t) = \alpha + (r_i^2 - \alpha)\exp[-\frac{2\mu}{k}t], ~~\theta = C\int^t_{t_i}\frac{dt}{r^2} 
\label{orbit}
\end{equation}
 putting $ \alpha = \frac{m\rho_0\kappa^2}{2k} $.  The orbit describes the spiral which starts with  the 
initial point $ r= r_i $ and converges to  the limiting radius: $ r  = \sqrt{\alpha} $.  
Then noting that the first integral in $ K_{cl} $ vanishes as a result of (\ref{instanton}),  we have
\begin{equation}
K_{cl}   =  \exp\big[-\frac{C^2 \mu}{kh}\int_{r_i}^{r_f} \frac{dr}{r(r^2 - \alpha )}\big].  
\end{equation}
Then we get  $ K_{cl}(r_f, r_i)  = \exp[ X] $ with 
\begin{equation} 
X =  \beta\big[\log\big(\frac{r_f^2 - \alpha}{r_i^2 - \alpha}\big)   - \log\big(\frac{r_f}{r_i}\big) \big] , 
\end{equation}
where we put $ \beta = \frac{C^2\mu}{2kh\alpha} $. 
Using the transition amplitude thus calculated, the probability is calculated to be 
\begin{eqnarray}
P(r_f) & = & \int_0^{\infty}  K_{cl}(r_f, r_i)P(r_i)dr_i 
\end{eqnarray}
and choosing the initial distribution $ P(r_i) = \delta (r_i - r_0) $, one gets 
\begin{equation}
P(r_f) = \big[\frac{r_0(r_f^2 - \alpha)}{r_f(r_0^2 - \alpha)} \big]^{\beta} . 
\end{equation}
This result indicates that the probability beyond the limiting radius vanishes.

\section{An aspect from the transport theory}

Up to now, our consideration of the generalized FP  equation has been restricted to the overdamping case, that is, $ \mu \gg  \bar\kappa $. Owing to this 
restriction, the standard FP equation is converted to the Schr\"odinger equation, by which we can use the resources of quantum mechanics.   Now there arises a  
problem: What  about the opposite case; namely,  the current $ {\bf j}_2 $ is dominant. That is, 
 the extremely  opposite limit  $ \mu \ll \bar\kappa $ holds together with the simultaneous restriction that $ h $ is enough small to be discarded, 
 then  (\ref{GFP}) turns out to be 
\begin{equation}
 \frac{\partial P }{ \partial  t } 
=  \bar\kappa  \sum_i\frac{\partial}{\partial {\bf R}_i} \cdot \{ ({\bf k}\times \frac{\partial H_{eff}}{\partial {\bf R}_i})P \} , 
\label{FP2}
 \end{equation} 
which is alternatively written as the Liouville equation 
\begin{equation}
\frac{\partial P }{\partial  t } = - \{  P, \bar\kappa H_{eff} \}, 
 \label{Liouville}
\end{equation}
where  $ \{....\}  $ denotes the Poisson bracket 
$$  \{A,B\} = \sum_i \big(\frac{\partial A}{\partial X_i}\frac{\partial B}{\partial Y_i}  - \frac{\partial A}{\partial Y_i}\frac{\partial B}{\partial X_i}\big) . 
$$
The righthand side of (\ref{Liouville})  is proportional to  $ \nabla \cdot (\sum_i {\bf v}_i P) $, which can be the divergence of the probability flow 
with the "phase space velocity" for each vortex: 
$$
{\bf v}_i (= \frac{d{\bf R}_i}{dt})  = {\bf k} \times \frac{\partial H_{eff}}{\partial {\bf R}_i} . 
$$
Thus (\ref{Liouville}) apparently suggests that there does not occur the equilibrium state in the vortex motion, whereas 
the gradient term in (\ref{GFP}) drives the equilibrium state.   In other words, the underdamping regime means that the dissipative 
as well as fluctuation force are rather weaker than the repulsive forces acting  between vortices. 
As a result of this,  there may not maintain the stable equilibrium state among the vortex gas.  
As such,  if the remaining term of (\ref{FP}) coming from $ {\bf j}_1 $, written as  $ G $, namely, 
 \begin{equation}
 G = \sum_i \big[\frac{h}{2} \frac{\partial^2P}{\partial {\bf R}_i^2} + \mu\frac{\partial}{\partial {\bf R}_i}  
 \cdot \frac{\partial H_{eff}}{\partial {\bf R}_i} P\big] , 
 \label{G}
 \end{equation}
 which can be treated as if  the "collision term", (\ref{Liouville}) 
can connect with the transport equation \cite{LP2}; 
 \begin{equation}
 \frac{\partial P }{\partial  t } - \{ \bar\kappa H_{eff}, P \} = G . 
 \label{inhomo}
 \end{equation}
 This can be treated by applying the perturbation scheme using an  iteration procedure. 
Let us write $ P = P_0 + P_1 $, where $  P_0 $ means the unperturbed term, which satisfies the stationary 
equation  $ \{H_{eff}, P\} = 0 $. By introducing the linear operator defined as 
\begin{equation}
\{ H_{eff}, P\} \equiv {\cal L}_HP , 
\end{equation}
then we have the equation for the perturbed term $ P_1 $, 
\begin{equation}
\frac{\partial P_1}{\partial t} -{\cal L}_H P_1  = G(P_0) 
\label{perturbed}
\end{equation}
and $ G(P_0) $ is the one for which the unperturbed solution $ P_0 $ is substituted in (\ref{G}). 
 The formal solution for (\ref{perturbed}) can  be obtained with the aid of the method of variation of parameters. First 
the homogenous equation is formally solved as 
\begin{equation}
P_1(t) = \exp[{\cal L}_H t] P_1(0) . 
\end{equation}
To look for a special solution of the inhomogeneous equation, we put $ P(t) = \exp[{\cal L}t] Q(t) $. 
By substituting this into (\ref{perturbed}) the equation for $ Q $ is derived as 
\begin{equation}
\frac{dQ}{dt} = \exp[-{\cal L}_Ht]G(P_0) , 
\end{equation}
from which we obtain a special solution  
\begin{equation}
P_s(t) = \exp[{\cal L}_Ht]\int \exp[-{\cal L}_Ht']G(t') dt' . 
\end{equation}
This serves as  a  formal perturbation solution for the transport equation. The more detailed analysis 
will be left for a future study.

\section{Concluding remarks}

The stochastic approach to the quantum vortex gas in two dimension has been 
investigated.  The starting point is the Hamiltonian dynamics for he vortex gas in which the coordinate $ (X_i, Y_i) $ 
form  a canonical pair each other. 
This can be transcribed to  the Langevin equation with the Gaussian white noise. Owing to the 
white noise, the Langevin equation is converted to the  functional integral, which results in the generalized Fokker-Planck(FP)  equation. 
In particular we have examined  the  overdamping limit yielding  the standard FP equation, for which 
we have examined several aspects in detail.   As for the underdamping case,  we have discussed it briefly, but 
there  may still remain a variety of problems  to be explored. The study will be left for future research. 
As a final remark, it would be interesting to address the problem to extend the 
 present Langevin and FP  formalism to the three dimensional dynamics that shows 
 an intricate process of entanglement of vortex curves \cite{Nemirovskii,Yui}.

\begin{appendix}

\section{Reduction to the path integral}

Here a prescription is given for the some step leading to the path integral form (\ref{PI}). 
Using the Fourier transform of the delta functional in (\ref{transition2}), it follows that 
\begin{align}
 K[{\bf R}(t)\vert {\bf R}(0)] &  = \int_{ {\bf{R}} (0) }^{ {\bf{R}} (t) } 
 \prod_i \exp \left[- \int^{t}_{0} \frac{{\bf \zeta}_i^2(t)}{2h}dt \right] \nonumber \\
  &  \times \prod_{i=1}^n \prod_t\exp[ 2\pi  i \int  \lambda_i(t) \{{\bf F}_i(t) - \zeta_i(t)\} dt] \nonumber \\
  & \times  \mathcal{D}{\bf F}_i(t)  \mathcal{D}[{\bf \zeta}_i(t)] \mathcal{D}[{\bf \lambda}_i(t)], 
\end{align}
then by  carrying out the Gaussian functional integral  over $ \zeta(t) $ and $ \lambda(t)  $, one gets 
\begin{equation}
 K[{\bf R}(t)\vert {\bf R}(0)] = \int \exp \left[ -\frac{1}{2h} \int_0^t \sum_i {\bf F}_i^2(t)dt \right] \prod_{i=1}^n \mathcal{D}{\bf F}_i . 
\end{equation}
This is converted to the functional integral over the vortex centers; 
\begin{eqnarray}
  K[{\bf R}(t)\vert {\bf R}(0)] & = & \int_{ {\bf{R}} (0) }^{ {\bf{R}} (t) }  \exp 
  \left[- \frac{1}{2h}\int_0^t  \sum_i \left(\frac{d{\bf R}_i}{dt} + {\bf A}_i \right)^2dt \right]  \nonumber \\
   & \times  &  \prod_i J({\bf R}_i) \mathcal{D}[{\bf R}_i] . 
  \label{PI2}
\end{eqnarray}
Here $ J({\bf{R}}) $ is the functional Jacobian given by
\begin{eqnarray}
  J({\bf{R}}) & = & {\rm{det}} \left( \frac{ \delta {\bf{F}}({\bf R}(t) )}{\delta{{\bf R}(t')}} \right) 
  \end{eqnarray}
  and after some steps of calculating the functional determinant \cite{TK}, this  leads to  
  \begin{equation}
 J =  \exp \left[ \int_{0}^{t} \frac{1}{2} \frac{ \partial }{ \partial {\bf{R}}} \cdot {\bf{A}} 
 dt \right] . 
 \end{equation}
By noting the exponential form, this factor can be incorporated into the action function and hence it plays 
a crucial role in determining the form of the FP equation given in the main text. \\

\section{ A brief sketch for N-particle problem}

We give an outline for treating the reduced quantum many particle system described by  the short range 
repulsive force coming from the inverse square interaction that balances with the attractive force coming from the harmonic potential. 
We adopt the procedure of the method of the collective coordinate \cite{tomonaga}.   The central idea is to separate the original particle degree of freedom 
into the collective degree and the internal one.  
In what follows we borrow it  with aiming at an application to 
the present problem.   According to Tomonaga, a natural candidate of the collective 
coordinate for the present case can  be chosen as  $ {\cal Q}_c  = \frac{1}{2}\sum_i (X_i^2 - Y_i^2)  $ 
together with the conjugate momentum
\begin{equation}
{\cal P}_c = \frac{1}{2}\sum_i (X_i\frac{\partial}{\partial X_i}  - Y_i\frac{\partial}{\partial Y_i}) . 
\end{equation}
The commutation relation for these becomes 
\begin{equation}
[ {\cal Q}_c, {\cal P}_c] = \frac{1}{NR_0^2}  \langle \sum_i (X_i^2 + Y_i^2) \rangle , 
\label{commutator}
\end{equation}
where $ \langle ....\rangle $ is an average appropriately defined and $ R_0 $, which is a circle radius, chosen such that (\ref{commutator}) satisfies the 
canonical commutation relation. The {\it quantum mechanical Hamiltonian}  $ {\cal H}_{eff} $ is thus written as a form of the coupling between 
the collective coordinate $ {\cal Q}_c, {\cal P}_c $ 
and the internal coordinate, say $ ({\cal Q}_{in}, {\cal P}_{in} ) $:  
\begin{align}
\hat {\cal H}_{FP}  & = {\cal H}_0({\cal Q}_{in}, {\cal P}_{in})  + {\cal H}_1({\cal Q}_{in}, {\cal P}_{in}){\cal Q}_c \nonumber \\
  & + {\cal H}_2({\cal Q}_{in}, {\cal P}_{in}){\cal Q}_c^2 + \frac{1}{2I}{\cal P}_c^2 . 
\end{align}

Having accomplished the separation of variables, the wave function can be expressed as the 
direct product: 
\begin{equation}
\tilde P = \psi_c({\cal Q}_c)\psi_{in}({\cal Q}_c, \{{\cal Q}_{in}, {\cal  P}_{in}\}) . 
\end{equation}
When the adiabatic separation is assumed, the internal coordinates are fixed and eliminated by 
integrating over them, namely,  $ {\cal H}_0, {\cal H}_1 $ and $ {\cal H}_2 $ are replaced by the 
expectation value with respect the internal state $ \psi_{in} $ which results in the wave function which 
is written in terms of the collective coordinate $ {\cal Q}_c $.  

As a special case, we consider three vortices of special configuration. Namely,  
we  suppose  the third and second vortices  are pinned at the origin: $ {\bf R}_3 = (0,0) and  {\bf R}_2 = (1, 0) $ respectively 
and hence the stationary counterpart of the eigenvalue equation  becomes  $ H_{eff}P = \epsilon P $ with the 
 "Hamiltonian" (\ref{inverseR^2})
\begin{align}
H_{eff} & =  H^0 + V ,   \nonumber \\
H^0 & =   - \frac{h}{2}\nabla_1^2 +\frac{B\mu}{ \mu^2 + \bar\kappa^2}R_1^2 + \frac{A\mu}{ \mu^2 + \bar\kappa^2 }\frac{1}{R_1^2} \nonumber \\
V  & =  \frac{ A\mu}{\mu^2 + \bar\kappa^2 }\frac{1}{(X-1)^2 + Y^2} . 
\end{align}
The crude estimate for this can  be carried out by applying the perturbation procedure if $ H^0 $ is regarded as non-perturbative term and 
treating $ V $ as the perturbation:  The solution for $ H^0 $ has been  obtained  for the two-vortices  case and the lowest energy change 
can be simply obtained by taking the expectation value with respect to the lowest eigenstate state of $ H^0 $.

\end{appendix}


\begin{thebibliography}{99}
\bibitem{foot1} This approximation is justified by noting that the localization of the vortex is much larger than the 
inter-vortex distance $ {\bf R}_{ij} $.
\bibitem{foot2} This comes from that the factor  $ \frac{1}{\bar\kappa^2 + \mu^2} $ appears in the right hand side of (\ref{langevin}).
\end{thebibliography}
\end{document}